\documentstyle[aps,prl,epsf,floats]{revtex}
\newcommand{\beq}{\begin{equation}}
\newcommand{\eeq}{\end{equation}}
\newcommand{\beqs}{\begin{eqnarray}}
\newcommand{\eeqs}{\end{eqnarray}}

\newcommand{\lsim}{\mathrel{\raisebox{-.6ex}{$\stackrel{\textstyle<}{\sim}$}}}

\draft

\tighten

\begin{document}

\twocolumn[\hsize\textwidth\columnwidth\hsize\csname
@twocolumnfalse\endcsname

\title{Quark Dipole Operators in Extended Technicolor Models}

\author{Thomas Appelquist$^1$, Maurizio Piai$^1$, and Robert Shrock$^2$}
\address{$^1$ Department of Physics, Sloane Laboratory, Yale University,
New Haven, CT 06520 \\ C. N. Yang Institute for Theoretical
Physics, State University of New York, Stony Brook, NY  11794}

\baselineskip 6.0mm

\tighten

\maketitle

\begin{abstract}

We study diagonal and transition quark dipole operators in a class of extended
technicolor (ETC) models, taking account of the multiscale nature of the ETC
gauge symmetry breaking and of the mixing among ETC interaction eigenstates.
Because of this mixing, terms involving the lowest ETC scale can play an
important role in dipole operators, and we focus on these terms.  We derive
from experiment new correlated constraints on the quark mixing angles and
phases. Our bounds yield information on mixing angles individually in the up-
and down-sectors, for both left- and right-handed quark fields and thus
constrain even quark mixing parameters that do not enter in the CKM matrix.
With phases of order unity, we conclude that these mixing angles are small,
constraining future ETC model building, but plausibly in the range suggested by
the size of the CKM elements. These values still allow substantial deviations
from the standard model predictions, in particular for several CP violating
quantities, including the asymmetries in $b \rightarrow s \gamma$ and $B_{d}
\rightarrow \phi K_S$, $Re(\epsilon^\prime/\epsilon)$, and the electric dipole
moments of the neutron and the ${}^{199}$Hg atom.

\end{abstract}


\vskip2.0pc]

\section{Introduction}

Extended technicolor (ETC) provides a framework \cite {ETC} for the
generation of fermion masses in theories of dynamical electroweak symmetry
breaking. In this paper we study diagonal and transition quark magnetic and
electric, and chromomagnetic and chromoelectric, dipole moments in a class
of ETC models \cite{at94}-\cite{ckm}, taking account of the multiscale
nature of the ETC gauge symmetry breaking, and mixing between ETC
interaction eigenstates to form mass eigenstates for the fermions and gauge
bosons.  The transition electric and magnetic dipole moments contribute to
processes such as $b \to s \gamma$. A CP-violating transition chromoelectric
moment contributes to $Re(\epsilon^\prime/\epsilon)$ in the kaon system, and
the diagonal electric and chromoelectric dipole moments contribute to a
nonzero electric dipole moment (EDM) for the neutron and atoms such as
${}^{199}$Hg. We work out the predictions for ETC contributions to these
quantities and compare these with measurements or bounds to obtain
constraints on the models.  We ask whether these constraints can be
satisfied without fine tuning in these models and find that they can. This
is an extension to quarks of our study of lepton dipole moments in Ref.
\cite{dml}.

The class of ETC models in Refs. \cite{at94}-\cite{ckm} is based on the
gauge group SU(5)$_{ETC}$ which commutes with the standard model (SM) gauge
group. It breaks sequentially to a residual exact SU(2)$_{TC}$ technicolor
gauge symmetry, naturally producing a hierarchy of charged lepton and quark
masses. Thus, SU(5)$_{ETC} \to$ SU(4)$_{ETC}$ at a scale $\Lambda_1$, with
the first-generation SM fermions separating from the others;
then SU(4)$_{ETC}
\to$ SU(3)$_{ETC}$ at a lower scale $\Lambda_2$ and SU(3)$_{ETC} \to$
SU(2)$_{TC}$ at a still lower scale $\Lambda_3$, with the second- and
third-generation fermions separating in the same way, leaving the
technifermions.

The models of Ref. \cite{ckm} exhibit charged-current flavor mixing,
intra-family mass splittings without excessive contributions to the
difference $\rho-1$ where $\rho=m_W^2/(m_Z^2 \cos^2\theta_W)$, a dynamical
origin of CP-violating phases in the quark and lepton sectors, and a
potential see-saw mechanism for light neutrinos without the presence of a
grand unified scale \cite{nt}. A key ingredient is the use of relatively
conjugate ETC representations for both down-quark and charged lepton fields
\cite{ckm}. The choice of SU(2) for the technicolor group (i) minimizes the
TC contributions to the electroweak $S$ parameter, (ii) with a
SM family of technifermions in the fundamental representation of
SU(2)$_{TC}$, can yield an approximate infrared fixed point \cite{wtc} and
associated walking behavior, and (iii) makes possible the mechanism for
light neutrinos.

The sequential breaking of the SU(5)$_{ETC}$ to SU(2)$_{TC}$ is driven by
the condensation of SM-singlet fermions which are part of the
models. At the scale $\Lambda_{TC}$, technifermion condensates break the
electroweak symmetry.  The models do not yet yield fully realistic fermion
masses and mixings, and they have a small number of unacceptable
Nambu-Goldstone bosons arising from spontaneously broken $U(1)$ global
symmetries. Additional interactions at energies not far above $\Lambda_1$
must be invoked to give them sufficiently large masses.

Nevertheless, the models share interesting generic features, including a
mechanism for CP violation, that are worth studying in their own right. We
approach this study phenomenologically, relying on only the generic features
and using current experimental data to constrain parameters and guide future
model building within the general class.

In any ETC model, the bilinear fermion condensates forming at each stage of
ETC breaking have nonzero phases, providing a natural, dynamical source of
CP violation.  Below the electroweak symmetry breaking scale, the effective
theory consists of the SM interactions, mass terms for the quarks, charged
leptons, and neutrinos, and a tower of higher-dimension operators generated
by the underlying ETC dynamics. Here we focus on the dimension-5 operators
describing the electric/magnetic and chromoelectric/chromomagnetic dipole
moments of the quarks. In a companion paper \cite{kt}, we discuss the impact
of dimension-6 operators.

\section{Quark mass matrices}

The dipole operators are related to the dimension-3 mass terms of the
up-type and down-type quarks, given in general by
\beq
{\cal L}_m = -\bar{f}_{L,j} M^{(f)}_{jk} f_{R,k} + h.c.
\label{m}
\eeq
where $f$ label the ETC eigenstates of the $Q=2/3$ and $Q=-1/3$ quarks,
respectively and the indices $j,k$ label generation number. The mass
matrices $M^{(f)}$ can in general be brought to real, positive diagonal form
$M^{(q)}$ by the bi-unitary transformation
\beq U^{(f)}_L M^{(f)} U^{(f) \ -1}_R = M^{(q)} \ . \label{biunitary} \eeq
Hence, the interaction eigenstates $f$ are mapped to mass eigenstates $q$
via
\beq f_\chi = U^{(f) \ -1}_\chi q_\chi \ , \quad \chi=L,R \label{ffm} \eeq
where $q=(u,c,t)$ and $q=(d,s,b)$ for $Q=2/3$ and $Q=-1/3$ respectively. In
this way, the Cabibbo-Kobayashi-Maskawa (CKM) quark mixing matrix entering the
charged current interactions is generated:
\beq
V = U_L^{(u)}\,U_L^{(d)\,\dagger} \,.
\label{v}
\eeq

Each of the matrices $U^{(f)}_\chi$, $\chi=L,R$, depends generally on three
angles $\theta^{(f)\chi}_{mn}$, $mn =12,13,23$, and six (independent) phases. 
Using the conventions of \cite{pdg}, we write 
\beqs
& &  U^{(f)}_\chi = \cr\cr
& & P^{(f)\chi}_\alpha R_{23}(\theta^{(f)\chi}_{23})
P^{(f)\chi \ *}_\delta R_{13}(\theta^{(f)\chi}_{13})P^{(f)\chi}_\delta
R(\theta^{(f)\chi}_{12})P^{(f)\chi}_\beta
\label{ufchi} \eeqs
where $R_{mn}(\theta^{(f)\chi}_{mn})$ is the rotation through
$\theta^{(f)\chi}_{mn}$ in the $mn$ subspace, $P^{(f)\chi}_\alpha$ and
$P^{(f)\chi}_\beta$ are given by
\beq P^{(f)\chi}_a = {\rm diag}( e^{ia^{(f)\chi}_1}, e^{ia^{(f)\chi}_2},
e^{ia^{(f)\chi}_3}), \quad a=\alpha,\beta \label{pfchi} \eeq
and $P^{(f)\chi}_\delta={\rm diag}(e^{i\delta^{(f)\chi}},1,1)$.  The mixing
angles are typically small if the off-diagonal $M^{(f)}_{jk}$'s are more
suppressed than the diagonal ones.

In ETC models, the off-diagonal entries of the quark mass matrices
$M^{(f)}$ arise via mixing among the ETC gauge bosons. In the model of Ref.
\cite{ckm}, employing a relatively conjugate ETC representation for the
down-type quarks, this is true also of the diagonal elements
\cite{ckmindex}. We note that in this model, $M^{(u)}$ is hermitian, so that
$U^{(u)}_L = U^{(u)}_R$, while $M^{(d)}$ is a more general complex matrix.

As for the phases, a complete theory should allow the computation of all the
observable ones \cite{lane_cpv}. In this paper, having neither a complete
theory nor arguments to suggest that the phases are small, we derive bounds
on combinations of mixing angles and phases. We then bound the mixing angles
with the assumption that the phases are generically of order unity. An
important future study will be to see whether this is naturally the case. We
comment further on this and the strong CP problem in
section~\ref{sec:strong}.

\section{Electromagnetic and Color Dipole Moment Matrices}

The magnetic and electric dipole-moment matrices $ D^{(f)}$ of the
quarks appear in the dimension-5 operators
\beq
{\cal L}_{DM}=\frac{1}{2} \bar f_L  D^{(f)} \sigma_{\mu\nu} f_R
F^{\mu\nu}_{em} + h.c.
\label{dipoleop}
\eeq
Similarly, the color (chromo-) magnetic and electric dipole-moment
matrices $ D^{(f)}_c$ enter the operators
\beq
{\cal L}_{CDM}=\frac{1}{2} \bar f_L T_a  D^{(f)}_c \sigma_{\mu\nu} f_R
G_a^{\mu\nu} + h.c.  ,
\label{colordipoleop}
\eeq
where $T_a$ and $G_a^{\mu\nu}$ denote a generator, and the field-strength
tensor, for color SU(3)$_c$.  (We include the color SU(3)$_c$ coupling $g_s$ in
our definition of $D^{(f)}_c$, just as it is usual to include the
electromagnetic coupling $e$ in the definition of the electric dipole moment;
we note that some authors separate $g_s$ out from their definition of
$D^{(f)}_c$.)

In the class of ETC models we consider, the ETC gauge bosons do not carry
SM quantum numbers. Hence, in the respective diagrams that
produce the dipole moment matrices and color dipole moment matrices, the
photon and gluon couple only to the virtual (techni)fermions. Therefore,
\beq
D^{(f)}_c = \frac{g_s}{eQ_f} D^{(f)}
\label{cdmrel}
\eeq
where $g_s=g_s(\mu)$ is evaluated at the appropriate scale $\mu$.

Transforming to the mass-eigenstate basis, we have
\beq
\bar f_L  D^{(f)} \sigma_{\mu\nu} f_R F^{\mu\nu}_{em} + h.c. = \bar q_L
D^{(q)} \sigma_{\mu\nu} q_R F^{\mu\nu}_{em} + h.c.,
\label{dipoleopuu}
\eeq
where
\beq
D^{(q)} = U^{(f)}_L  D^{(f)} U^{(f) \ -1}_R  \ .
\label{ddtilderel}
\eeq
Analogously, $D^{(q)}_c = U^{(f)}_L D^{(f)}_c U^{(f) \ -1}_R$. Both
$D^{(q)}$ and $D^{(q)}_c$ are independent of $P^{(f)\chi}_\beta$,
$\chi=L,R$.

Decomposing $D^{(q)}$ into hermitian and anti-hermitian parts, $D^{(q)} =
D^{(q)}_H + D^{(q)}_{AH}$, where $D^{(q)}_{H,AH} = (1/2)(D^{(q)} \pm D^{(q)
\ \dagger})$, the dipole operator takes the form $(1/2)[\bar q D^{(q)}_H
\sigma_{\mu\nu} q + \bar q D^{(q)}_{AH} \sigma_{\mu\nu} \gamma_5 q ]
F^{\mu\nu}_{em}$. Then the EDM of $q_j$ is
\beq
d_{q_j} = -iD^{(q)}_{AH,jj} \ .
\label{df}
\eeq
Defining $D_{c,AH}^{(q)}$ analogously, the chromo-EDM of $q_j$ is
\beq
d_{c,q_j} = -iD^{(q)}_{c,AH,jj} \ .
\label{dfc}
\eeq

We have described previously \cite{ckm,dml} how the mass matrix $M^{(f)}$
and dipole matrix $\tilde D^{(f)}$ are estimated from an underlying ETC
theory and have noted how they are related in the presence of the mixing of
ETC interaction eigenstates to form mass eigenstates of the fermions and
gauge bosons. An important result that we need from that analysis is the
relation
\beq
 D^{(f)}_{jk} \simeq \frac{e M^{(f)}_{jk}}{\Lambda_{jk}^2}
\label{dmrel}
\eeq
where each $\Lambda_{jk}$ is a dimensionful parameter of order the scale above
which the ($j \leftrightarrow k$) ETC propagator becomes soft. This structure
reflects the fact that the leading dipole contribution contains two additional
inverse factors of the ETC scale(s) relative to the mass, and that the
corresponding integral is again sensitive to physics at the ETC scales.
$\Lambda_{jk}$ is no greater than ${\rm min}(\Lambda_j, \Lambda_k)$, and can be
less. The fact that its $(j,k)$ dependence is nontrivial implies that
$D^{(f)}_{jk}$ is not, in general, $\propto M^{(f)}_{jk}$. It is therefore not
diagonalized by the transformation that diagonalizes $M^{(f)}$; this
transformation yields, instead, a non-diagonal and complex form for the dipole
matrix $D^{(q)}$ of Eq. (\ref{ddtilderel}). Thus mixing has an important effect
on quark dipole moments in ETC models.

For numerical estimates of the dipole matrix, we take the ETC breaking scales
to be $\Lambda_1 \simeq 10^3$ TeV, $\Lambda_2 \simeq 10^2$ TeV, and $\Lambda_3
\simeq 4$ TeV, as in our previous work.  Since the ETC interactions are strong
at these scales, there is resultant uncertainty in the calculations; this is 
understood in our bounds.  We focus on the contribution to each element of
the dipole matrix in $D^{(q)}$ in the mass-diagonal basis of third-family
physics arising at the lowest ETC scale $\Lambda_3$. Then we have
\beq
D^{(q)}_{jk} \simeq \frac{e Q_q \ m_{q_3} F^{(f)}_{jk,3}}{\Lambda_3^2}
\label{djk}
\eeq
where $m_{q_3}=m_t,m_b$ for the $u,d$ sectors respectively, and where
$F^{(f)}_{jk,3}$ is a dimensionless function of the parameters in
$U^{(f)}_L$ and $U^{(f)}_R$, of $O(1)$ for generic values of these
parameters. Terms involving exchange of heavier ETC vector bosons with
masses $\Lambda_1$ and $\Lambda_2$ are also present but are suppressed by
the propagator mass ratios $\Lambda_3^2/\Lambda_j^2$, $j=1,2$.  Part of this
propagator suppression may be compensated for by the property that these
other terms involve fewer small mixing angle factors, and hence they are not
necessarily negligible; however, the $\Lambda_3$-scale terms on which we
focus should provide a rough measure of the overall ETC contributions.

The experimental constraints will demand small mixing angles, so we record
here the small-angle form of the function $F^{(f)}_{jk,3}$. Using Eq.
(\ref{ddtilderel}) and Eq. (\ref{ufchi}), we have
\beq F^{(f)}_{33,3} \simeq 1 + ... \ , \label{f333} \eeq
\beq F^{(f)}_{23,3} \simeq
e^{i(\alpha_2^{(f)L}-\alpha_3^{(f)L})}\theta_{23}^{(f)L} + ... \ ,
\label{f233} \eeq
\beq F^{(f)}_{32,3} \simeq
e^{i(-\alpha_2^{(f)R}+\alpha_3^{(f)R})}\theta_{23}^{(f)R} + ...\ ,
\label{f323} \eeq
and, for $j,k \ne 3$,
\beq F^{(f)}_{jk,3} \simeq \eta_{jk} e^{i[(\alpha_j^{(f)L}-\alpha_3^{(f)L})-
(\alpha_k^{(f)R}-\alpha_3^{(f)R})]}\theta^{(f)L}_{j3} \theta_{k3}^{(f)R} +
...\ ,
\label{fjk3} \eeq
where each expression is accurate up to a real coefficient of order unity,
$\eta_{jk}$ can contain $\delta^{(f)\chi}$ phases, and
... denote higher order terms.

\section{The Strong CP Problem}
\label{sec:strong}

Before considering the phenomenology of the dimension-5 operators, with
their CP violating phases, we discuss briefly the strong CP problem within
the class of ETC models being considered. Can these models lead to the
necessary condition
\beq |\bar\theta| \lsim 10^{-10} \, \label{thetabarlimit} \eeq
where
\beq
\bar\theta = \theta -[arg(det(M^{(u)})) + arg(det(M^{(d)}))] \ ,
\label{thetabar}
\eeq
with $\theta$ appearing via the topological term
\beq
\frac{\theta g_s^2}{32 \pi^2} G_{a \ \mu\nu} \tilde G_a^{\mu\nu}
\label{ggdualterm}
\eeq
in the QCD Lagrangian?

The quark mass matrices $M^{(f)}$ ($f = u,d$) in the effective theory below
$\Lambda_{TC}$ are generated by integrating out short-distance physics at
scales ranging from $\Lambda_{TC}$ to the highest ETC scale $\Lambda_1$.
Above $\Lambda_{1}$, all fermions are massless. Some of the global chiral
symmetries are anomalous, and hence are broken by instantons.  The $F_{\mu
\nu} \tilde{F}^{\mu\nu}$ terms associated with each (nonabelian) gauge
interaction may be rotated away by chiral transformations through the
relevant global anomalies. In particular, this renders $\theta=0$ for
SU(3)$_c$ in the underlying theory. In the effective low energy theory,
then, we have $\bar\theta = -arg(det(M^{(u)})) - arg(det(M^{(d)}))$. If
$\bar \theta \neq 0$, the rotation (2) to the real diagonal mass basis will,
of course, regenerate the topological term through the anomaly. In the
models of Refs. \cite{at94}-\cite{ckm}, $M^{(u)}$ is hermitian, so
$\bar\theta$ resides in $M^{(d)}$.

More generally, the condition $|arg(det(M^{(u)})) + arg(det(M^{(d)}))| \lsim
10^{-10}$ can be rewritten by letting
\beq
U^{(f)}_\chi = e^{i\phi^{(f)}_\chi} {\cal U}^{(f)}_\chi \in {\rm U}(3)
\label{uufactorization}
\eeq
where ${\cal U}^{(f)}_\chi \in {\rm SU}(3)$ and $\chi = L,R$.  Then
$det(U^{(f)}_\chi)=e^{i\phi^{(f)}_\chi}$, and, from Eq. (\ref{biunitary}),
\beq
det(M^{(f)})=e^{i(-\phi^{(f)}_L+\phi^{(f)}_R)}det(M^{(f)}_{diag.}) \ .
\label{detmf}
\eeq
Hence, the necessary condition reads
\beq |\sum_{f=u,d} (-\phi^{(f)}_L+\phi^{(f)}_R)| \lsim 10^{-10} \ .
\label{argdetmud} \eeq
Other phases, entering the CKM matrix or the dipole operators, enter through
the unimodular matrices ${\cal U}^{(f)}_\chi$, and are, in this sense, distinct
from the strong CP phase.

In the notation of Eq. (\ref{ufchi}),
\beq det(U^{(f)}_\chi)=exp[i\sum_j (\alpha_j^{(f) \chi} + \beta_j^{(f)
\chi})] \ , \quad \chi=L,R \label{detuf} \eeq
and, in terms of these quantities, Eq. (\ref{argdetmud}) reads
\beqs \biggl | \sum_{f=u,d} \sum_j \biggl [
-(\alpha_j^{(f)L}+\beta_j^{(f)L})\nonumber\\+
(\alpha_j^{(f)R}+\beta_j^{(f)R}) \biggr ] \biggr | \lsim 10^{-10} \ .
\label{argdetmudalpha} \eeqs
(independent of $\delta^{(f)\chi}$). So in terms of these parameters, only one
linear combination (a sum over flavors $j$) of the CP-violating phases
$\alpha^{(f)\chi}_j$ and $\beta^{(f)\chi}_j$ is tightly constrained.

Whether a resolution of the strong CP problem will emerge in the class of
models considered here is not yet clear \cite{lane_cpv}. These models include
certain Nambu-Goldstone bosons, one of which can be associated with a
Peccei-Quinn dynamical relaxation of $\bar \theta$ to zero. But, as noted
above, these Nambu-Goldstone bosons must be given large masses by new
interactions, eliminating this approach to solving the problem. Whatever the
resolution of the strong CP problem turns out to be, an important point is that
the relevant phase combination, entering in Eq. (\ref{argdetmud}) or
equivalently Eq.  (\ref{argdetmudalpha}), involves a sum over the generational
phases (labelled by $j$), whereas other CP-violating phase combinations which
contribute to the quantities considered here (cf. Eqs.
(\ref{f233})-(\ref{fjk3})) involve differences of generational phases (and
$\delta^{(f)\chi}$).  One must, therefore, analyze the effects of these
flavor-dependent phases, as we do here.

\section{Off-Diagonal Dipole Moments}

In this section, we begin our phenomenological discussion by focusing on the
off-diagonal entries in the matrices $D^{(q)}$ and $D^{(q)}_c$. We turn to
the diagonal elements in Section V. The off-diagonal elements, both the
CP-conserving and CP-violating pieces, contribute to transitions $q \to
q^\prime \gamma$ and $q \to q^\prime g$, where $g$ denotes a gluon (which
hadronizes). We derive constraints on ETC contributions to the resultant
hadron decays.  These complement our previous study of upper limits from the
electromagnetic leptonic decays $\mu \to e \gamma$, $\tau \to \mu \gamma$,
and $\tau \to e \gamma$ in Ref. \cite{dml}.

\subsection{$b \to s \gamma$ and $ b \to s g$}

We first consider the processes $b \to s \gamma$ and $b \to s g$. The former
underlies the decays $B \to X_s \gamma$, where $X_s$ denotes a
semi-inclusive final state containing an $s$ quark.   For this purpose, one
constructs an effective Hamiltonian describing the physics at energies below
the electroweak scale by integrating out the heavy $W$ and $Z$ gauge bosons
and the top quark. QCD effects are then taken into account through the
renormalization group (RG) running of the Wilson coefficients of this
effective Hamiltonian down to the scale relevant for the physical process of
interest, where the matrix elements of the operators are computed.  We use
the operator basis $O_k$ as in \cite{bsgthy}. The relevant effective
Hamiltonian for the processes $b \to s \gamma$ and $b \to s g$, keeping
dominant terms, is
\beqs & & {\cal H} \simeq -\frac{G_F}{\sqrt{2}}\,V_{ts}^* V_{tb}\, \biggl
[ \sum_{k=1}^{10}\,C_k\,O_k \cr\cr & + & C_{7\gamma}\,O_{7\gamma} +
C_{8g}\,O_{8g} \biggr ] + h.c. \label{effham} \eeqs
where $V_{tb}$ and $V_{ts}$ are CKM matrix elements. The $O_k$, $k=1,..,10$ are
dimension-6 four-fermion operators, while the last two operators are the
dimension-5 operators of primary interest in this paper,
\beq
O_{7\gamma} = \frac{e\, m_b}{4\pi^2} \, [\bar s_L \sigma_{\mu\nu} b_R] \,
F^{\mu\nu}_{em}
\label{O7gamma}
\eeq
and
\beq
O_{8g} = \frac{g_s \,m_b}{4\pi^2} \, [\bar s_L \sigma_{\mu\nu} T_a b_R] \,
G^{\mu\nu}_a \ .
\label{O8g}
\eeq
QCD running between the electroweak scale, taken here as $m_Z$, and the
scale of $B$-decays $\mu_b$ introduces mixing among the Wilson coefficients, in
such a way that observables determined by any particular operator $O_k$ at
the low scale depend on a combination of the Wilson coefficients at the
electroweak scale.

The Wilson coefficients of the effective Hamiltonian receive contributions
from SM physics as well as ETC interactions. Our focus is on the ETC
contributions to the dipole operators and we thus take the first 10 Wilson
coefficients, computed at the common scale ${\cal O}(m_Z)$, to be determined
by the SM interactions only: $C_k = C_k^{SM}$ for $k=1,...,10$. For the
dipole-operator coefficients $C_{7\gamma}$ and $C_{8g}$ at the electroweak
scale, we have
\beq
C_{7\gamma,8g} = C_{7\gamma,8g}^{SM} + \Delta C_{7\gamma,8g}
\eeq
where the increments are due to the ETC interactions and are given by
\beq \Delta C_{7\gamma} \simeq -\frac{2\sqrt{2}\pi^2}{G_F\Lambda_3^2}
\frac{Q_f\,F^{(d)}_{23,3}}{V_{ts}^* V_{tb}} \label{dc7} \eeq
and
\beq \Delta C_{8g} \simeq -\frac{2\sqrt{2}\pi^2}{G_F\Lambda_3^2}
\frac{F^{(d)}_{23,3}}{V_{ts}^* V_{tb}} \ . \eeq

After input of the SM contributions at the electroweak scale and QCD
evolution to $\mu_b$, the Wilson coefficients can be written in the form
\cite{bsgthy},
\beq C_{7 \gamma}(\mu_b) \simeq -0.3 + 0.7 \Delta  C_{7\gamma}+0.09 \Delta
C_{8g} \ , \label{c7eff} \eeq
\beq C_{8g}(\mu_b) \simeq -0.15 + 0.7  \Delta C_{8g} \ . \label{c8eff} \eeq
The first term in each case is the SM contribution, computed using input CKM
parameters based on global fits \cite{bsgthy,parodi}. Each lies essentially
along the real axis, because the rephasing-invariant quantity $V_{cs}^*
V_{cb}/(V_{ts}^* V_{tb})$ has a negligibly small complex phase. In the case of
$C_{7 \gamma}(\mu_b)$, the dominant ETC correction comes from the $0.7 \Delta
C_{7\gamma}$ term.

The branching ratio $BR(B \to X_{s} \gamma)$ is proportional to $|C_{7
\gamma}(\mu_b)|^2$. Experimentally, $BR(B \to X_s \gamma) = (3.34 \pm 0.38)
\times 10^{-4}$ \cite{nakao}, in agreement with the SM value at the $10 \%$
level. This leads to an allowed annular region in the complex $C_{7
\gamma}(\mu_b)$-plane, a band of width $\pm 5\%$ relative to the SM value.
The resultant constraint on the magnitude of the ETC contribution depends on
its CP-violating phase. This phase depends in turn on the phase differences
in the small-angle expressions Eqs. (\ref{f233}) and (\ref{f323}), which
enter $\Delta C_{7\gamma}$ through Eq. (\ref{dc7}).

Two possibilities suggest themselves. One is that the ETC contribution to
$C_{7\gamma}(\mu_b)$ is less than about $5\%$ in magnitude relative to the
SM value. There is then no constraint on the value of the phase,
and we find
\beq |\theta^{(d)\chi}_{23}| \lsim 0.02 \quad {\rm for} \quad \chi=L,R .
 \label{bsglimit} \eeq
Another possibility is that the magnitude of the ETC contribution is larger,
but that the phase is such as to yield a value for $C_{7 \gamma}(\mu_b)$
within the allowed annular region in the complex plane. The latter case
involves some correlation between the magnitude and phase of the ETC
contribution, but would allow a larger value for the mixing angles
$\theta^{(d)\chi}_{23}$, with $\chi=L,R$.

To explore further the possibility of a significant relative phase between
the standard model and ETC contributions to the $b \to s \gamma$ amplitude,
we consider CP-violating asymmetries in $B$ decay. Let $A_{CP}(i \to f)$
denote the CP-violating asymmetry in the rates for an initial particle $i$
to decay to a final state $f$: $A_{CP}(i \to f) = (\Gamma_{i \to f} -
\Gamma_{\bar i \to \bar f})/ (\Gamma_{i \to f} + \Gamma_{\bar i \to \bar
f})$.  Current data yields $-0.093 \le A_{CP}( B \to X_{s} \gamma) \le
0.096$ (Belle, \cite{nakao,belle_bsgasym}) and $-0.06 \le A_{CP}(B \to X_{s}
\gamma) \le 0.11$ (BABAR,\cite{babar_bsgasym}), both at the 90 \% CL.  These
limits are consistent with the standard model, in which, using global fits
to CKM parameters, this asymmetry  is predicted to be $\simeq 0.005$
\cite{acpbsg}. This agreement constrains ETC contributions.  We use the
expression (e.g. \cite{bsgthy,acpbsg})
\beqs
A_{CP}( B \to X_{s} \gamma) & \simeq &
\frac{1}{|C_{7\gamma}(\mu_{b})|^2} \biggl [ a_{27}Im(C_2(\mu_{b})
C_{7\gamma}^*(\mu_{b})) \cr\cr & + &
a_{87}Im(C_{8g}(\mu_{b})C_{7\gamma}^*(\mu_{b})) \biggr ]
\eeqs
where the coefficients $a_{ij}$ are known quantities, and where we have kept
only the largest interference terms. The Wilson coefficient $C_2$
corresponds to the operator $O_2 = 4 [\bar s_L \gamma_\mu c_L][\bar c_L
\gamma^\mu b_L]$, and is present already at tree level in the standard model.
The ETC contribution may be computed using Eqs. (\ref{c7eff}),
(\ref{c8eff}), and (\ref{effham}). The current bounds are such that
$|Im(F^{(d)}_{jk,3})| \lsim 0.1$ for $jk=32,23$.

Hence, neglecting the special possibility of maximal destructive
interference between the SM and ETC contributions to the $b \to s \gamma$
amplitude, the measured rate and the bounds on the CP-violating asymmetry
for this decay together constrain $|\theta_{23}^{(d)\chi}| \lsim O(0.02 -
0.1)$. That these mixing angles might not be too far below this range is
suggested by the fact that combinations of $\theta^{(d)L}_{jk}$ and
$\theta^{(u)L}_{jk}$ enter the CKM mixing matrix (\ref{v}). In general, one
might expect that the individual mixing angles $\theta^{(u)\chi}_{jk}$ and
$\theta^{(d)\chi}_{jk}$ would be comparable to the corresponding measured
CKM angle $\theta_{jk}$ in $V$. The measured value of the CKM angle
$\theta_{23}$ is $\theta_{23} \simeq 0.04$, in roughly the same range.

We remark briefly on other CP-violating observables affected by the possible
ETC modification of $C_{7\gamma}$ and $C_{8g}$. In the standard model,
time-dependent CP-violating asymmetries in the decays $B_d^0,\bar{B}_d^0
\rightarrow \phi K_S$, arising from one-loop (penguin) diagrams, are
predicted to be the same as those in the tree-level decays
$B_d^0,\bar{B}_d^0 \rightarrow J/\psi\,K_S$. ETC contributions to the $\phi
K_S$ asymmetries through $C_{8g}$ could change this. At present, Belle and
BABAR measurements of the CP-violating asymmetries in $B_d^0,\bar{B}_d^0
\rightarrow \phi K_S$ disagree with each other \cite{pdg}, and hence it is
not clear if this data will confirm or deviate from the SM predictions.
Constraints from other gluon-penguin dominated processes such as
$B_d^0,\bar{B}_d^0 \rightarrow \eta^{\prime} K_S$ should also be taken into
account in a complete analysis.

\subsection{Other Off-diagonal Terms}

We next consider the process $s \to d \gamma$ which gives rise to radiative
hyperon and meson decays.  The relevant Hamiltonian is similar to that of
Eq. (\ref{effham}), with the respective replacements of $b$ by $s$ and $s$ by
$d$. Branching ratios for radiative hyperon decays are of order $10^{-3}$; for
example, $BR(\Sigma^+ \to p \gamma) = (1.23 \pm 0.05) \times 10^{-3}$ and
$BR(\Lambda \to n \gamma) = (1.75 \pm 0.15) \times 10^{-3}$ \cite{pdg}.
Angular asymmetries in decays of polarized hyperons have also been measured.  A
typical radiative $K$ decay is $K^+ \to \pi^+ \pi^0 \gamma$, with $BR(K^+ \to
\pi^+ \pi^0 \gamma) = (2.75 \pm 0.15) \times 10^{-4}$. Although the branching
ratios and asymmetries are only approximately calculable, owing to
long-distance contributions, the standard model yields an acceptable fit. The relevant
ETC-induced transition dipole moment is given by Eq. (\ref{djk}) with $jk=12$
and $jk=21$. We find that the ETC contributions are safely smaller than the SM
one and hence these decay rates and asymmetries do not yield interesting
constraints on the fermion mixing.

The process $ s \to d g$ provides a tighter constraint.  Its
chromomagnetic and chromoelectric dipole elements produce a (virtual) gluon
which then hadronizes. Most importantly, the ETC-induced transition
chromo-EDM provides a new contribution to direct CP violation in $K_L \to
2\pi$ decays. The latter is measured by the quantity
$Re(\epsilon^\prime/\epsilon)$, which is determined via
\beq
\biggl | \frac{\eta_{+-}}{\eta_{00}} \biggr |^2
\simeq 1 + 6 Re \biggl ( \frac{\epsilon^\prime}
{\epsilon} \biggr )
\label{etaratio}
\eeq
where $\eta_{+-}$ and $\eta_{00}$ are given in terms of measured amplitude
ratios by $A(K_L \to \pi^+\pi^-)/A(K_S \to \pi^+\pi^-) = \eta_{+-} \simeq
\epsilon + \epsilon^\prime$ and $A(K_L \to \pi^0\pi^0)/A(K_S \to \pi^0\pi^0)
= \eta_{00} \simeq \epsilon - 2\epsilon^\prime$.  Experimentally,
$Re(\epsilon^\prime/\epsilon) = (1.8 \pm 0.4) \times 10^{-3}$ \cite{pdg}.
There are uncertainties in theoretical estimates of
$Re(\epsilon^\prime/\epsilon)$ in the standard model owing to difficulties
in calculating the relevant matrix elements and in choosing input values of
some parameters such as the strange quark mass \cite{bertolini}.
Nevertheless, we can deduce a rough bound from the requirement that the
contribution from the ETC-induced transition quark chromo-EDM operator not
be excessively large.  We obtain
\beq
|Im(F^{(d)}_{12,3})| \lsim 10^{-3} - 10^{-4} \ .
\label{eppboundF}
\eeq

This bound can be satisfied with the phase differences in $Im(F_{12,3})$ of
order unity providing that
\beq |\theta^{(d)\chi}_{13}\,\theta^{(d)\chi^{\prime}}_{23}| \lsim 10^{-3} -
10^{-4} \ ,
\label{eppboundtheta} \eeq
where $\chi, \chi' = L , R $ or $R , L$.

A similar limit on the corresponding up-type angles emerges from the process
$c \to u \gamma$. It leads to radiative decays such as $D^+ \to \pi^+ \pi^0
\gamma$, $D^+ \to \rho^+\gamma$, $D^0 \to \pi^+ \pi^- \gamma$, and $D^0 \to
\rho^0 \gamma$. Using the limit $BR(D^0 \to \rho^0 \gamma) < 2.4 \times
10^{-4}$ \cite{pdg} that has been established on one of these possible
decays, we derive the bound
\beq |\theta_{13}^{(u)\chi}\theta_{23}^{(u)\chi}| \lsim 0.008 \
\label{t1323u} \eeq
where $\chi = L, R$.

The possibility that these products of angles are not too far below the
bounds (\ref{eppboundtheta}) and (\ref{t1323u}) is suggested by the fact
that the mixing angles $\theta^{(d)L}_{jk}$ and $\theta^{(u)L}_{jk}$ enter
the CKM quark mixing matrix (\ref{v}). As noted earlier, one might expect
that $\theta^{(u)\chi}_{jk}$ and $\theta^{(d)\chi}_{jk}$ would be comparable
to the corresponding measured CKM angle $\theta_{jk}$ in $V$. The measured
value of the CKM angle $\theta_{13}$ is $\theta_{13} \simeq 0.004$, in
roughly the same range.

The ETC-induced transition dipole moments also contribute to the elementary
decays $b\rightarrow d \gamma$, affecting both CP-conserving and
CP-violating observables.  However, the experimental search for these
decays, both via inclusive modes and such exclusive modes as $B \to \rho
\gamma$ and $B \to \omega \gamma$, is difficult because of the small
branching ratios and large backgrounds.  Current upper limits on branching
ratios for these exclusive modes from Belle and BABAR are $\sim 10^{-6}$
\cite{nakao}.  These do not yield significant constraints on ETC
contributions. In the up-sector, because of the large top mass and limited
data on top decays, there are only weak limits on the radiative modes $t \to
u \gamma$ and $t \to c \gamma$.

\subsection{Discussion}

The study of the off-diagonal elements of the matrices $D^{(q)}$ and
$D_{c}^{(q)}$, focusing on physics at the lowest ETC scale $\Lambda_3$,
leads to constraints that can be satisfied with CP-violating phase
differences of order unity and reasonable limits on the relevant mixing
angles. These limits are consistent with values in a range suggested by the
fact that the angles $\theta^{(d)L}_{jk}$ and $\theta^{(u)L}_{jk}$ determine
the measured CKM angles through Eq (\ref{v}).

It is worth observing that at least one combination of the phases
$\alpha^{{(f)}L}$, $\beta^{{(f)}L}$, and $\delta^{(f)L}$ for $f=u,d$
must be of order unity. The CKM rephasing-invariant product
\beq J =(1/8)\sin(2\theta_{12}) \sin(2\theta_{23})\sin(2\theta_{13})\cos
\theta_{13} \sin\delta  \eeq
is quite small, $|J| \sim 10^{-5}$, but this suppression arises from the
small CP-conserving mixing angles. The intrinsic CP-violating phase angle
$\delta$ defined in the standard parametrization by $V_{ub}=e^{-i\delta}
\sin\theta_{13}$, is not
 small.  Indeed, current CKM fits give, for the ratio
$\bar\eta/\bar\rho$, which is equal to $\tan\delta$ in the standard model, a
value $\bar\eta/\bar\rho \simeq 2$ \cite{parodi}.

\section{Diagonal Electric and Chromoelectric Dipole Moments}

The diagonal electric and chromo-electric dipole moments for the up and down
quarks arising from ETC interactions provide tighter constraints on certain
combinations of phase differences and mixing angles. These moments derive from the
diagonal elements of $D^{(q)}_{jk}$ (Eq. \ref{djk}). Recall that this
expression is correct to leading order in ETC scales, depending explicitly
on the inverse square of only the lowest ETC scale $\Lambda_3$. Thus,
\beq
d_u = \frac{e \ Q_u}{g_s} d_{c,u} \simeq \frac{e \ Q_u \  m_t \
Im(F_{11,3}^{(u)})}{\Lambda_3^2}
\label{du}
\eeq
\beq d_d = \frac{e \ Q_d}{g_s} d_{c,d} \simeq \frac{e \ Q_d \ m_b \
Im(F_{11,3}^{(d)})}{\Lambda_3^2} \ ,
\label{dd} \eeq
with the small-angle expressions for the $F_{11,3}^{(f)}$ given by Eq.
(\ref{fjk3}) and involving flavor-differences of phases.

We stress that these mixing-induced terms can be the dominant contributions to
the up- and down-quark EDM's only if the sum over all flavor-phases (in Eqs.
(\ref{argdetmud}) or (\ref{argdetmudalpha})) is negligibly small, that is if
the strong CP problem has been solved. We note that the pure electroweak
contribution to quark electric dipole moments has been estimated to be $\lsim
10^{-32}$ e-cm and hence {\it is} negligibly small \cite{rev,dnfuture}.

CP-violating electric dipole moments such as those of the neutron and
certain atoms like ${}^{199}$Hg receive contributions from the quark EDM's
and the quark color EDM's. The latter lead to CP-violation in the hadronic
wave function (the CEDM enters as a correction to $t$-channel gluon exchange
between the bound quarks). There are also contributions from the
CP-violating triple-gluon operator $c_{abc}G_{a \ \lambda\mu} \tilde
G_b^{\mu\nu}G^{\lambda}_{c \ \nu}$ \cite{weinberg_ggg}, and loop-induced
CP-violating $W^+W^- \gamma$ vertices \cite{tgv}.

We focus here on the quark-EDM contributiuons. To estimate them, one starts as
in the case of the off-diagonal elements with operators defined at short
distances, uses renormalization-group methods to evolve these to hadronic
distance scales $\sim 1$ fm, and then computes the relevant hadronic matrix
elements. For two reasons, we adopt a simpler approach here.  Because the
direct SM contribution to the EDM's is many orders of magnitude smaller than
the expected ETC contribution, we neglect the SM contribution here. And because
the computation of the hadronic matrix elements, involving only
first-generation quarks, is more uncertain than for the off-diagonal matrix
elements, we also neglect the RG running of the ETC contribution.
                                                                               
\subsection{Bounds from EDM's}

The current experimental upper bound on the neutron electric dipole moment
$d_n$ is $|d_n| < 6.3 \times 10^{-26}$ e-cm \cite{dnILL}. In setting
constraints, we assume that there are no accidental cancellations between
different contributions to the experimentally observable EDM's.  For an
estimate of the hadronic matrix element of the quark EDM operators, $\langle
n | \bar f \sigma_{\mu\nu} \gamma_5 f | n \rangle$, $f=u,d$, various methods
yield roughly similar results, which are comparable to the static quark
model relation $d_n = (1/3)(4d_d-d_u)$. Using these estimates, we infer from
the above limit on $|d_n|$ that $|Im(F^{(u)}_{11,3})| \lsim 1 \times
10^{-6}$ and $|Im(F^{(d)}_{11,3})| \lsim 3 \times 10^{-5}$.

The same quantities enter into the quark color EDM's and can, in principle,
be bounded from their contributions to $d_n$. Since QCD is nonperturbative
at the low energies relevant here, there are uncertainties in the
proportionality factors connecting the CEDM's $d_{c,f}$, $f=u,d$, to $d_n$
(e.g., \cite{falk}), and there is the related question of what value to use
for the color gauge coupling $g_s$ in Eq. (\ref{cdmrel}) at such low
energies.  In general, from these CEDM's one obtains limits that are
comparable to the ones above on $|Im(F^{(f)}_{11,3})|$, $f=u,d$.

The most stringent limits on these quantities are obtained from upper bounds
on EDM's of atoms, in particular from ${}^{199}$Hg. Experimentally,
$|d_{{}^{199} Hg}| < 2.1 \times 10^{-28}$ e-cm \cite{hg199}, which is about
a factor of 50 smaller than the upper limit on $|d_n|$. From this we obtain
the bounds
\beq
|Im(F^{(u)}_{11,3})| \lsim 0.3 \times 10^{-7}
\label{imf113limit_u}
\eeq
\beq
|Im(F^{(d)}_{11,3})| \lsim 0.6 \times 10^{-6} \ .
\label{imf113limit_d}
\eeq
We note that in the class of models Refs. \cite{at94}-\cite{ckm},
$Im(F^{(u)}_{11,3})$ (\ref{fjk3}) vanishes identically. This is because
$M^{(u)}$ is hermitian in these models and therefore $U^{(u)}_L =
U^{(u)}_R$.

 From Eq. (\ref{fjk3}), we see that Eqs. (\ref{imf113limit_u}) and
(\ref{imf113limit_d}) constrain a product of $\theta^{(f)L}_{13}
\theta_{13}^{(f)R}$ times the imaginary part of a phase factor, for $f =
u,d$. If the phase differences are of order unity, then
\beq |\theta_{13}^{(d)L}\theta_{13}^{(d)R}| \lsim 0.6 \times 10^{-6} \ ,
\label{t1313d} \eeq
with a tighter bound on $|\theta_{13}^{(u)L}\theta_{13}^{(u)R}|$ if
$M^{(u)}$ is not hermitian. The bound (\ref{t1313d}) is comparable to that
on the corresponding product of charged-lepton mixing angles coming from the
current limit on the electron EDM \cite{dml}.

The above bound may be satisfied with $|\theta_{13}^{(d)L}| \simeq
|\theta_{13}^{(d)R}| \leq 0.0008$. We note again that the corresponding
angle $\theta_{13}$ in the CKM matrix $V$ has the measured value $\simeq
0.004$, only a factor of five larger. Furthermore, the expression for
$\theta_{13}$ contains terms proportional to products such as
$\theta_{12}^{(u)L} \theta_{23}^{(d)L}$ which could be the dominant
contribution and naturally be of order $0.004$. Hence, although the bound
(\ref{t1313d}) and its analogue for $f=u$ do imply quite small values for
the indicated products of rotation angles, they can plausibly be satisfied
in ETC models that successfully predict the CKM matrix.

We note, by contrast, that if all the above mixing angles are of order the CKM
angle $\theta_{13} \simeq 0.004$, then the relevant combination of phase
differences in the down sector must be rather small, $\lsim 0.04$. The
corresponding limit on the up-sector would be even smaller, of order $0.002$,
but would be automatically satisfied in models where $M^{(u)}$ is hermitian.

Finally, we comment that an $s$-quark (chromo-) EDM can also contribute to
the ${}^{199}$Hg EDM. This contribution is more difficult to compute since
there are no valence $s$ quarks in the nucleon. It can be roughly estimated,
leading to a bound on $|Im(F^{(d)}_{22,3})|$. This bound is much weaker than
the above bound on $|Im(F^{(d)}_{11,3})|$, but it could have important
implications for the mixing angles $\theta_{23}^{(d)L}$ and
$\theta_{23}^{(d)R}$.

\subsection{Discussion}

The quark EDM's and chromo-EDM's correspond to the diagonal elements of the
matrices $D^{(q)}$ and $D_{c}^{(q)}$. The bounds on these quantities, in
particular from the measured limit on the EDM of ${}^{199}$Hg, lead to tight
constraints on mixing angles and/or CP-violating phase differences in the
individual subsectors $q=u,d$.  These derive from the contributions to the
elements of $D^{(q)}$ and $D_{c}^{(q)}$ from physics at the lowest ETC
scales $\Lambda_3$ (e.g. Eq. (\ref{djk})). If the phase differences are of
order unity, then the down-type mixing angles are bounded as in Eq.
(\ref{t1313d}), with an even tighter bound in the up sector if $M^{(u)}$ is
not hermitian. The bound (\ref{t1313d}) is comparable to that on the
corresponding product of charged-lepton mixing angles coming from the
current limit on the electron EDM \cite{dml}.

\section{Conclusions}

We have shown that in the class of ETC models \cite{at94}-\cite{ckm}, mixing
effects significantly affect predictions for diagonal and transition
magnetic and electric dipole and color dipole moments.  We have used
measurements of $b \to s \gamma$, $c \rightarrow u \gamma$ and
$Re(\epsilon^\prime/\epsilon)$, and limits on CP-violating electric and
chromoelectric dipole moments of quarks to set new constraints on the mixing
angles and phases in the unitary transformations from the quark flavor
(ETC-interaction) eigenstates to the mass eigenstates. The analysis focuses
on physics at the lowest ETC scale $\Lambda_3$, taken to be a few TeV to
allow for the measured value of $m_t$. While there are ETC contributions to
these processes involving all of the ETC scales, the terms arising at
$\Lambda_3$ should provide a rough measure of the overall ETC effect.

Our bounds provide new information about quark mixings since they apply
separately to the individual charge sectors $q=u$ and $q=d$. By contrast,
the measured CP-conserving angles and CP-violating phase in the CKM matrix
arise from the (mismatch of the) unitary transformations $U^{(u)}_L$ and
$U^{(d)}_L$. Our bounds also constrain the unitary transformations
$U^{(q)}_R$, $q=u,d$, which do not enter in the charged weak current and
associated CKM matrix.

We make two observations about the bounds we have derived, assuming the
CP-violating phase differences to be of order unity. First of all, each of
the mixing angles in the diagonalization of the quark mass matrices is
relatively small, a restrictive requirement for the further development of
ETC models, perhaps along the lines of Refs. \cite{at94}-\cite{ckm}. (The
same is true for the charged leptons \cite{dml}.) Secondly, while the mixing
angles must be small, the bounds allow them to be "reasonable", that is, in
the same range as the measured values of the CKM angles (which are
expressible as combinations of mixing angles for left-handed quarks).

In the class of models of Refs. \cite{at94}-\cite{ckm}, small mixing angles
can emerge for the up-type quark masses because the off-diagonal terms
require mixing among the ETC gauge bosons, which is suppressed by ratios of
ETC scales, while the diagonal terms do not. The mechanism employed in this
class of models for the suppression of down-type quark masses and
charged-lepton masses requires ETC gauge-boson mixing to generate diagonal
as well as off-diagonal elements. Thus the ingredients for small mixing
angles in the down-quark sector are not yet as evident.

Finally, we note that the experimental constraints on the dimension-five
dipole operators still allow ETC models to produce sizable departures from
the SM prediction, in particular for several CP violating observables. These
include the EDM's of the neutron and of $^{199}$Hg, the direct CP violation
in the kaon system (measured as $Re(\epsilon^{\prime}/\epsilon)$) and the CP
asymmetry in the inclusive decay $b\rightarrow s \gamma$, or in the decay
$B_d \rightarrow \phi K_S$. Improvements in the experimental sensitivity and
reductions in the QCD uncertainties of the SM prediction could set more
stringent bounds or even allow detection of ETC effects.

\acknowledgements

We thank K.~Lane for stimulating discussions and M.~Frigerio for a
comment. This research was partially supported by the grants DE-FG02-92ER-4074
(T.A., M.P.) and NSF-PHY-00-98527 (R.S.).


\begin{thebibliography}{99}
\bibitem{ckmindex}
We note that in this model, the ETC mixing arises from the coupling to
certain condensing SM-singlet fermions, and takes the form of two-, three-,
and four-point ETC gauge-boson vertices.
\end{thebibliography}
\end{document}